\documentclass{emulateapj}

\shortauthors{INOUE and CHIBA}
\shorttitle{Three-dimensional Mapping of CDM Substructure}

\newcommand{\x}{{\mathbf{x}}}
\newcommand{\y}{{\mathbf{y}}}

\newcommand{\f}{\frac}
\newcommand{\T}{\tilde}

\newcommand{\bb}{\bibitem}
\newcommand{\BF}{\begin{figure}\begin{center}}
\newcommand{\EF}{\end{center}\end{figure}}
\newcommand{\BE}{\begin{equation}}
\newcommand{\EE}{\end{equation}}
\newcommand{\BEA}{\begin{eqnarray}}
\newcommand{\EEA}{\end{eqnarray}}
\newcommand{\ti}{\textit}

\newcommand{\ms}{M_{\odot}}

\newcommand\bu{{\bf 1}}
\newcommand\bg{{\bf \Gamma}}

\begin{document}


\title{Three-dimensional Mapping of CDM Substructure at Submillimeter Wavelengths}

\author{Kaiki Taro Inoue  \altaffilmark{1} and Masashi Chiba \altaffilmark{2}}
\altaffiltext{1}{Department of Science and Engineering, 
Kinki University, Higashi-Osaka, 577-8502, Japan}
\altaffiltext{2}{Astronomical Institute, Tohoku University,
Sendai 980-8578, Japan}

\begin{abstract}

The cold dark matter (CDM) structure formation model predicts
that about 5-10 \% of a typical galactic 
halo of mass $\sim 10^{12} \ms$
is in substructures with masses $\lesssim 10^8 \ms$.
To directly detect such substructures, we propose to 
observe dust continuum emission from a strongly lensed 
QSO-host galaxy using a large submillimeter 
interferometer array with a high angular 
resolution of $\sim 0.01$arcsec such as the planned 
Atacama Large Submillimeter Array (ALMA). 
To assess their observational feasibility, 
we numerically simulate millilensing of 
an extended circular 
source by a CDM substructure modeled as a
tidally truncated singular isothermal sphere
(SIS) embedded in a typical QSO-galaxy lens system, B1422+231, 
modeled as a singular isothermal ellipsoid (SIE) with an 
external constant shear and a constant convergence. 
Assuming an angular resolution of $0.01$arcsec, we find that the 
angular positions of $\sim 10^8 \ms$ substructures at several kpc
from the center of the macrolens halo can be 
directly measured if the 
size of the dust continuum emission region and
the gradient of the surface brightness  
at the position of the perturber
are sufficiently large.
From the astrometric shift on a scale of a few times $10~$mas 
of an image perturbed by a subhalo 
with respect to an unperturbed macrolensed image, 
we can break the degeneracy between subhalo mass and distance   
provided that macrolensing parameters are determined 
from positions and fluxes of multiple images.
\end{abstract}

\section{Introduction}

Recent high-resolution $N$-body simulations
of the cold dark matter (CDM) structure formation model
show an excess in the number of subhalos (``substructures'')
in a galaxy-sized halo in comparison with the number of Milky Way 
or M31 satellites (Klypin et al. 1999; Moore et al. 1999).
If the CDM-based structure formation scenario is the correct one,
then we expect a large number of invisible satellites inside a halo
of galaxies.

Currently, gravitational lensing is the 
only probe that can test the abundance of 
these invisible satellites or equivalently, substructures in a
galaxy-sized halo. In fact, gravitationally multiply lensed QSOs 
have recently been used for putting limits on the surface 
density and the mass of such
invisible substructures (Mao \& Schneider 1998; Metcalf \& Madau
2001; Chiba 2002; Metcalf \& Zhao 2002; Dalal \& Kochanek 2002; 
Bradac et al. 2002). Statistical analyses on these quadruple QSO-galaxy
lensing systems that show an anomalous flux ratio 
imply the presence of substructures along
the line-of-sight to the multiple images (Metcalf \& Zhao 2002; 
Evans \& Witt 2003; Keeton, et al. 2003).

However, the location of a subhalo   
along the line-of-sight is poorly constrained. 
In fact, more massive extragalactic $10^{8}-10^9\ms $ halos along
the line-of-sight to the multiple images may also significantly 
contribute to altering the flux ratio of multiple images (Metcalf
2005). In order to determine the 
distances to these subhalos, we need to have
more information in addition to the position 
and flux of multiple images (Yonehara, et al. 2003). 

To directly map out the three-dimensional locations of 
$10^8 \ms$ invisible subhalos, 
we propose to observe the
dust continuum emission from a strongly lensed 
QSO-host galaxy using a large submillimeter 
interferometer array with a high angular 
resolution of $\sim 0.01$ arcsec such as the ALMA.

Recent submillimeter observations revealed that about 60 \% 
of strongly lensed active galactic nuclei(AGNs) 
are luminous at a submillimeter 
wavelength of $850\mu$m (Barvainis \& Ivison 2002) 
implying that a large amount of 
energy emitted by stars and AGNs in the QSO host galaxy
has been absorbed by dust grains and
then re-radiated at longer wavelengths. 
The dust is heated to 
temperatures of 20-50K and radiates as a modified blackbody at
far-infrared wavelengths (Wiklind 2003). At a redshift $z\simeq 3-4$, 
the corresponding energy peak is at a submillimeter wavelength 
$\sim 0.1$ mm. The region responsible for 
this dust continuum emission has a typical
scale of $10^2$pc to a few kpc. Therefore, the effects of 
an extended source with an angular size larger than the Einstein radius
of a perturber will be easily observed at 
submillimeter wavelengths (Blain 1999; 
Wiklind \& Alloin 2002).

In this paper, we show our simulation results   
of millilensing of an extended circular source 
by a CDM substructure whose Einstein radius is smaller than 
the source size. We also show that the astrometric shifts of images
perturbed by a subhalo can reveal the three-dimensional position of 
a subhalo along the line-of-sight. As an example, we choose 
a typical QSO-galaxy lensing system, 
B1422+231, with cusp-caustics in order to check 
the observational feasibility of mapping substructures.

\section{Simulation}

The typical QSO-galaxy lensing system 
B1422+231 shows an anomalous flux ratio.
The images consist of three highly magnified images 
A, B, and C, and a faint one D located near the 
lens galaxy. The observed mid-infrared flux ratios 
are $A/|B|=0.94\pm 0.05$ and $C/|B|=0.57\pm 0.06$ 
(Chiba et al. 2005). The radio and 
optical flux ratios are consistent with 
the mid-infrared values within a $\sim 10$ \% error.
If we assume that the gravitational potential 
of the macrolens is a smooth one, we expect a cusp-caustic relation
in the flux ratio, $f\equiv (A+B+C)/(|A|+|B|+|C|)=0$. 
However, B1422+231 shows 
a 20 \% deviation from the $f=0$ relation. Assuming 
that such an anomaly is caused by 
the presence of an SIS perturber in the macrolens halo, 
the anomaly might be in image A (Dobler \& Keeton 2005).
The redshifts of the source and the lens are
$z_S=3.62$ and $z_L=0.34$, respectively. In what follows, we assume 
the following cosmological parameters: the present 
density of matter $\Omega_m=0.3$, 
the present density of the cosmological constant 
$\Omega_\Lambda=0.7$, and the Hubble parameter $h=0.7$, which yield
the angular diameter distances to the lens and to the source,
$D_L=1.00$~Gpc and $D_S=1.49$~Gpc, respectively.

To model the macrolens system, we adopt
an SIE in an external shear field in which the
isopotential curves in the projected surface perpendicular to the
line-of-sight are ellipses (Kormann et al. 1994).
For further details, see Chiba (2002). 
In what follows, we assume an angular 
resolution of $0.01$ arcsec which will be achieved by 
the planned ALMA at 850 $\mu$m.
Because the Einstein radius 
for an SIS with a one-dimensional velocity
dispersion $\sigma$ at $D_L=1.00$Gpc and $D_S=1.49$Gpc
is
\BE
\theta_{Ep}=1.0\times 10  \biggl (  
\f {  \sigma  } {21~\textrm{km~s}^{-1}  }     
\biggr )^2 \textrm{mas},
\EE
observation with an angular resolution of $0.01$arcsec
will reveal subhalos with a one-dimensional 
velocity dispersion $\sigma \gtrsim 20$ km~s$^{-1}$ for B1422+231.

As a model of a CDM subhalo, we consider a spherically symmetric,
tidally truncated SIS with one-dimensional velocity dispersion $\sigma$.
At a distance $r$ from the center of a macrolens galactic halo with 
a one-dimensional velocity dispersion $\sigma_0$, the tidal radius
is approximately given by $r_t \approx  r 
\sigma / \sigma_0 $. For an SIS with a one-dimensional 
velocity dispersion $\sigma \sim 20$ km~s$^{-1}$, 
the tidally truncated mass is 
$M_{\textrm{SIS}}\sim 10^8 \ms$.
Provided that the mass function for subhalos satisfies
$dn/dM \propto M^{-2}$, as many $N$-body simulations
suggest (e.g., Klypin et al. 1999), the differential lensing cross 
section $d S$
per logarithmic mass interval is 
$d S = \theta_{E_p}^2 d n/d \ln M \propto M^{1/3}$ 
because the area inside an Einstein radius for a tidally cut SIS 
is $\theta_{E_p}^2 \propto M_{\textrm{SIS}}^{4/3}$.
Thus, the contribution of massive subhalos to the substructure lensing
is somewhat significant in comparison with that of less massive ones.
We also assume that a perturbing subhalo is located at a distance
equal to the size of the Einstein radius of the macrolens halo
$r_E=D_L \theta_E=3.8$kpc from the macrolens center
(where $\theta_E$ is the Einstein angular radius of the macrolens halo).
This assumption can be verified as follows. Consider a
column with a unit section centered at a point $P$ aligned to 
the line-of-sight direction centered at a distance equal to the size  
of the Einstein radius of the macrolens $r_E$ from the macrolens center. 
We introduce a column distance $L$ from $P$ 
as a coordinate in the direction of the line-of-sight in the column.
If the subhalo distance is larger than the size of the Einstein radius
of the macrolens halo, $r \gg r_E$,  the tidally truncated mass
approximately satisfies $M_{\textrm{SIS}} \propto L$, since $r\sim L$. 
Therefore, 
the probability of having a subhalo with a fixed one-dimensional 
velocity dispersion $\sigma$  
per logarithmic column distance $\ln{L}$ is 
$d p(L)\sim L^{-1} d \ln{L}$,  
provided that the mass function 
for subhalos in the column 
with a fixed one-dimensional velocity dispersion  
satisfies $dn/dM \propto M^{-2}$. 
Thus, the lensing probability is expected 
to be larger for a smaller column distance $L$.

If the ``real'' mass function takes a different form, or the above
assumption is incorrect, the perturbing subhalo may 
reside at a position with a larger 
column distance $L>r_E$. Then the corresponding 
subhalo mass will be increased by a factor of $\sim L/r_E$. 

In our simulation, we put a
tidally truncated SIS with a mass of 
$2\times 10^{8}\ms$ near the A image.\footnote{We choose 
an appropriate position of the SIS that approximately 
gives the observed anomaly flux ratio, but we did not fix the
relative position of the images in a small source size limit.
Therefore, our simulation should be interpreted as a demonstration of 
extended source size effects rather than that of a 
best-fit model.}
The corresponding Einstein radius of the SIS 
is $\theta_{Ep}\sim 17$ mas.
As a model of emission from cold dusts, 
we assume a circularly symmetric Gaussian source
with a standard deviation $L=2.5\times 10^2$pc.

\section{Result}
The left panels in Figure 1 show lensed images 
without any perturbers, whereas the right panels 
show those perturbed by
an SIS perturber. As shown in a 
zoomed in picture of image A ({\it{top right}}),
the angular position of an SIS can be clearly identified by 
a dipole structure that consists of a pair
of dark and bright spots at the place where the surface brightness
gradient is non-vanishing (Inoue \& Chiba 2003). 
One can also notice that
shifts of an image outside the Einstein radius $\theta_{Ep}$ of an SIS 
with respect to the unperturbed macrolensed image are significantly 
suppressed ({\it{top right})}.

\begin{figure*}
\centerline{\includegraphics[width=16.5cm]{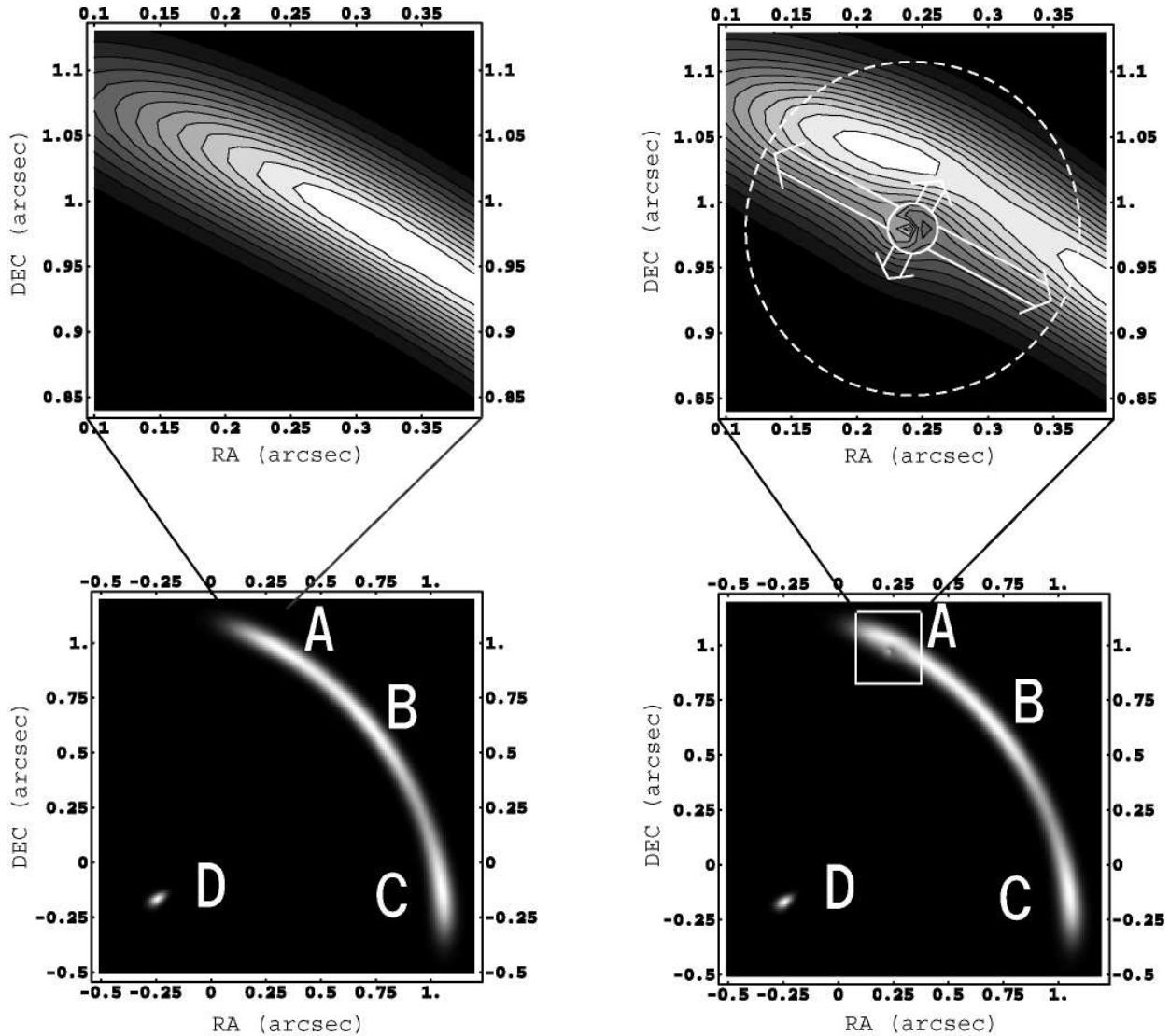}}
\label{image1}
\caption{Simulated images of B1422+231 at 
submillimeter wavelengths
 without any perturbers (\ti{left}) and with an SIS perturber 
(\ti{right})
for a Gaussian circular source with 
a standard deviation $L=2.5\times 10^2$~pc. 
A tidally truncated SIS with a mass of 
$2\times 10^{8}\ms$ is put at the center of a 
thick circle (\ti{top right}) near the A image.  The 
dashed circle represents a circle with the tidal radius 
centered at an SIS perturber. The angular resolution is
assumed to be 0.01 arcsec. Astrometric shifts in the tangential
direction and those in the radial direction in 
the coordinates aligned to the shear 
are represented by arrows.}
\end{figure*}
If we can approximate the lens system near the perturber 
locally as an SIS plus a constant shear $\gamma$ 
and a constant convergence $\kappa$, then the lens equation 
normalized by the Einstein radius $\theta_{Ep}$ of an SIS 
in the coordinates aligned to the shear can be 
written in terms of a source position $\y=(y_1,y_2)$ 
and an image position $\x=(x_1, x_2)$ as 
\BE
\y=(\bu-\bg)\x-\f{\x}{|\x|}, \label{eq:lens}
\EE
where 
\BE
\bg=
\pmatrix{
\kappa+\gamma  &0 \cr 0&\kappa-\gamma}.
\EE
Let $\x_0=(x_{01},x_{02})$ be the image position
for a macrolens model without any perturbers. 
In terms of the polar coordinates $(R_p,\phi)$ defined as
$\x=(R_p \cos \phi, R_p \sin \phi)$, the lens 
equation (\ref{eq:lens}) yields 
\BE
x_{01}^2(R_p-\xi_{2}^{-1})^2+x_{02}^2(R_p-\xi_{1}^{-1})^2=
(R_p-\xi_{1}^{-1})^2 (R_p-\xi_{2}^{-1})^2, \label{eq:fourth}
\EE
where $\xi_1\equiv 1-\kappa-\gamma$ and $\xi_2 \equiv 1-\kappa+\gamma $.
The astrometric shift $\Delta \x \equiv \x-\x_{0}$
as a function of $\x_0$ can be obtained by 
solving the fourth-order equation 
(\ref{eq:fourth}) with equation (\ref{eq:lens}) analytically. 
For a positive parity case $\xi_1>0, \xi_2>0$, an image at 
a horizontal axis $(x_{01},0)$ is shifted 
to $(x_{01}+\xi_{1}^{-1},0)$, whereas an image at 
a vertical axis $(0,x_{02})$ is 
shifted to $(0,x_{02}+\xi_{2}^{-1})$ (Inoue \& Chiba 2004).
One can see in the top right panel in 
figure 1 that image A, with 
positive parity $\kappa\sim 0.38, \gamma\sim 0.47$, 
is largely stretched in the tangential 
direction by $\xi_1^{-1}\sim 6.7$ and slightly stretched
in the radial direction by $\xi_2^{-1} \sim 0.92$
with respect to the center of the perturber.
The astrometric shift is significantly 
suppressed if the lensed image is placed outside the 
tidal radius centered at the SIS perturber, because the
projected gravitational potential decreases faster, as $R^{-1}$, 
than an SIS without truncation, where $R$ is the 
projected distance from the center of the SIS. 
If the redshift and the one-dimensional velocity dispersion 
$\sigma_0 $ of the macrolens halo are known, then we obtain
the tidal radius as a function of the distance 
from the center of the 
macrolens halo. From the observed angular size for which the
astrometric shifts are significantly suppressed,  
we can determine the distance from the center of the lens halo,
in principle.

Next, we show that a measurement of astrometric shifts can break 
the degeneracy between substructure mass and distance in the
line-of-sight to the image.
Consider a halo at redshift $z=z_h$ and 
a clump (a substructure) in the foreground of the halo at 
redshift $z_c<z_h$. Let $D_{s}$, $D_{ch}$, $D_{cs}$, and $D_h$ 
be the angular diameter distances to the source, 
between the clump and halo, between the clump and source, and 
to the halo, respectively. Provided that the angle
between the perturber and the macrolensed
image is sufficiently small in comparison with the Einstein
radius of the macrolens $\theta_E$, the effective 
lens equation for an SIS 
plus a constant shear and a constant 
convergence can be written as (Keeton 2003)
\BE
\T{\y}=(\bu-\T{\bg})\x-\f{\x}{|\x|}, 
\label{eq:eff-lens}
\EE
where $\T{\y}=(\bu-\beta \bg)^{-1} \y$, $\T{\bg}=
\bu-(\bu-\beta \bg)^{-1}(\bu-\bg)$, and 
$\beta=D_{ch}D_s/D_{cs}D_h$.
Because an 
effective unperturbed image position $\T{\x}_0$ that satisfies
$\T{\y}=(\bu-\T{\bg})\T{\x}_0$ is equal to the 
unperturbed macrolensed image position $\x_0$, 
astrometric shifts can be written as $\Delta \x=\x-\T{\x}_0$.
The effective convergence and shear in $\T{\bg}$ 
are then (Keeton 2003)
\BE
\kappa_{\textrm{eff}}=\f{(1-\beta)
(\kappa-\beta(\kappa^2-\gamma^2))}{(1-\beta \kappa)^2-\beta^2
\gamma^2},\gamma_{\textrm{eff}}=\f{(1-\beta)\gamma}
{(1-\beta \kappa)^2-\beta^2
\gamma^2}.
\EE
Thus, astrometric shifts are described in the same manner as
in the case when $\beta=0$. A tangential shift is $\Delta x_1=
(1-\kappa_{\textrm{eff}}-
\gamma_{\textrm{eff}})^{-1}$, and a radial shift is 
$\Delta x_2=
(1-\kappa_{\textrm{eff}}+
\gamma_{\textrm{eff}})^{-1}$ (Inoue \& Chiba 2004). 
As shown in figure 2, 
for images with positive parity, the 
tangential shift decreases as 
the distance between a perturber and a macrolens halo 
increases.
As the distance to the clump $D_c$ decreases, 
the astrometric shift becomes less conspicuous and more isotropic.
In fact, in the limit $D_c \rightarrow 0$ or, equivalently, 
$\beta \rightarrow 1$, 
the astrometric shifts converge to a perfect isotropic shift 
$\Delta \x = \x_0/|\x_0|$. 
Thus, the distance between a halo and a clump
can be measured by the absolute value and the anisotropy  
of the shifts, provided that the 
distances to the macrolens halo $D_h$ and the source $D_s$ 
are known already.

Similarly, we can calculate astrometric shifts 
$\Delta \x$ in the background case $z_c>z_h$. 
Again, the effective lens equation takes the same form
as in equation (\ref{eq:eff-lens}) (Keeton 2003),
\BE
\y=(\bu-\T{\bg})\T{\x}-\f{\T{\x}}{|\T{\x}|}, 
\label{eq:eff-lens2}
\EE
where $\T{\x}=(\bu-\beta \bg)\x$, $\T{\bg}=
\bu-(\bu-\bg)(\bu-\beta \bg)^{-1}$, and 
$\beta=D_{ch}D_s/D_{c}D_{sh}$ 
($D_{sh}$ is the angular diameter distance between the source
and the halo). We can define an effective  
position of the unperturbed image $\T{\x}_0$ satisfying
$\y=(\bu-\T{\bg})\T{\x}_0$, which leads to 
$\T{\x}_0=(\bu-\beta \bg)\x_0$.
Then astrometric shifts can be expressed as $\Delta \x=(\bu-\beta
\bg)^{-1} \Delta \T{\x} ((\bu-\beta \bg)\x_0)$, 
where $\Delta \T{\x}\equiv \T{\x}-\T{\x_0}$.
Thus, astrometric shifts in the background case 
are different from those in the foreground case 
with the same $\beta$, although the effective lens equation 
takes the same form in both cases. From the right panels 
in Figure 2 and 3, one can see that the shifts of images
near the horizontal axis $(x_1,0)$ 
are larger for the background case
in comparison with the foreground case with the same $\beta$. 
As shown in the left panel in Figure 2 and both panels in 
Figure 3, the astrometric shifts of images 
on the horizontal axis $\Delta \x= (\xi_1^{-1},0)$
and those on the vertical axis  
$\Delta \x=(0,\xi_2^{-1})$ do not 
depend on $\beta$ in the background case.
However, astrometric shifts of all the other
images do depend on $\beta$. As $\beta$ increases from 0,
astrometric shifts tend to become isotropic except for
images on the axes of coordinates aligned to the shear, but they 
cannot be perfectly isotropic as in the foreground case 
even if $\beta=1$. 

Now let us assess our assumption for constancy in the 
convergence $\kappa$ and shear $\gamma$. The absolute errors 
in astrometric shifts in the horizontal direction (aligned to the shear)
and in the vertical direction are written in terms of 
absolute errors in the convergence $\delta \kappa $ and shear $\delta
\gamma$ as $\delta(\Delta x_1) \sim \xi_1^{-1}( 
|\delta\kappa/\kappa|+|\delta \gamma/\gamma |)$ and 
$\delta(\Delta x_2) \sim \xi_2^{-1}( 
|\delta\kappa/\kappa|+|\delta \gamma/\gamma |)$, respectively.
In order to achieve a three-dimensional mapping with an angular resolution   
of a unit Einstein radius ($\sim \theta_{E_p}=1$), 
the errors $\delta \kappa $ and $\delta \gamma$ should satisfy  
$|\xi_i^{-1}|(|\delta\kappa/\kappa|+|\delta \gamma/\gamma |)\ll 1  
(i=1,2)$ at the scale of a unit Einstein radius.
In order to check this condition, we compute $\kappa$ and $\gamma$ 
in the neighborhood of an SIS perturber for our SIE model.
As shown in Figure 4, the variances in $\kappa$ and $\gamma$ 
at the scale of the perturber's Einstein radius $\theta_{E_p}=17$ mas  
are only of the order of a few \%. Thus, our approximation 
is verified for an SIS with one-dimensional dispersion $\sigma \sim 
20$ km~s$^{-1}$. In more general settings,  
the magnitude of spatial variation in the convergence and shear
at the scale of the Einstein radius $\theta_{E_p}$ of 
a perturber can be estimated 
as $\delta \kappa /\kappa \sim \delta \gamma/\gamma
\sim \theta_{E_p}/\theta_E$. If the macrolens is also an SIS
with a one-dimensional velocity dispersion $\sigma_0$, then 
we should have $\xi_i (\sigma/\sigma_0)^2 \ll 1$. In our model, 
this condition is roughly satisfied, 
because $\sigma_0 \sim 180$ km~s$^{-1}$.  
However, for subhalos with a much larger velocity dispersion $\sigma$, 
we should take the effect of spatial variance in the 
convergence $\kappa$ and shear $\gamma$ into account.

\section{Observational feasibility}
The observed values of the flux density at 850$ \mu$m of 
QSOs at $z=3-4$ with a starburst are typically several tens of mJy,
which implies an emission from dust with a temperature of 
30-60K from a fairly extended region of $10^{2}-10^{3}$~pc (Barvainis 
\& Ivison 2002). 
Let us assume that the far-infrared dust emission can be 
represented by a single-temperature
graybody emission with a rest-frame spectral energy distribution
(SED) (Wiklind 2003)
\BE
S(\nu_r,T_d)=\Omega  B(\nu_r,T_d) (1-\exp(-(\nu_r/\nu_0 )^
 \beta )),
\EE
 where $B(\nu_r,T_d)$ is the Planck function in terms of
a rest-frame frequency $\nu_r$ and a dust temperature
$T_d$, $ \beta =1-2$ is the model parameter that
controls the grain emissivity, $\nu_0$ is the critical 
frequency that controls the opacity of the dust, and $\Omega $
is the solid angle of the source seen by an observer.
A rest-frame SED $S$ is related to the observer-frame SED 
as $\tilde{S}(\nu)= S((1+z)\nu,T_d)/(1+z)$.
Assuming a model with $ \beta=1.5$, $\nu_0$=
6 THz, a dust temperature $T_d=$40 K, and 
a dust emission region from a circular 
disk with a radius $L=250$~pc, at 850 $\mu$m,  
the observed energy density would be $ \sim 2.6$ mJy. 

Because the ratio of the area $D$
within the Einstein radius of a $2\times 10^8\ms$ 
SIS perturber to the area of image A is approximately
$1/50$, the observed energy flux from the area $D$ will be 
$\sim 4\times10^{-4}$~Jy, taking 
a magnification factor of $\sim$6 into account.
In order to measure the 
difference in the energy flux at the scale of 
the Einstein radius of the SIS perturber, for a 
source with a smooth surface brightness distribution 
(e.g., Gaussian) one needs a sensitivity of 
at least $\sim 4\times10^{-5}$Jy. 
On the other hand, the sensitivity of ALMA
at 850 $\mu$m for a point source (signal-to-noise ratio S/N$=1$) 
is $1.2\times10^{-5}\times 
(\textrm{integration time}/5\textrm{hour})^{-1/2}$Jy. 
\footnote{See the ALMA home page at 
\\
http://www.eso.org/projects/alma/science/bin/sensitivity.html.}
Thus, ALMA has a sufficient sensitivity for a source
with a spatially varying surface brightness to 
detect the dipole structure caused by a $2\times 10^8 \ms$ 
SIS perturber. The required integration time would be
several hours. Intrinsic surface brightness variations 
in the source might make the identification of the clear dipole structure
rather difficult. However, such a difficulty can be avoided by 
comparing multiple macrolensed images, because intrinsic variations 
in one macrolensed image can be linearly mapped to those 
in another one.
\begin{figure*}
\centerline{\includegraphics[width=8cm]{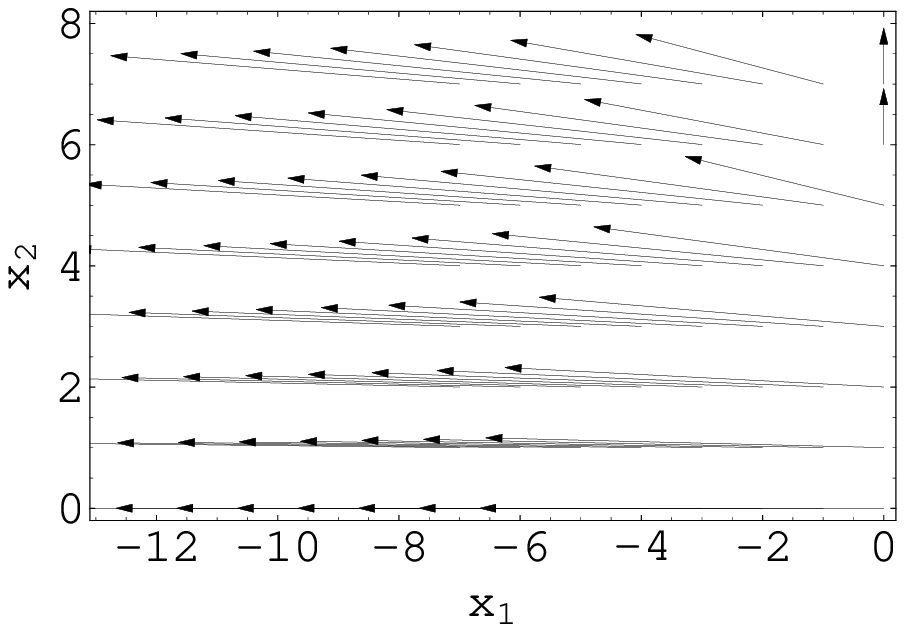}
            \includegraphics[width=8cm]{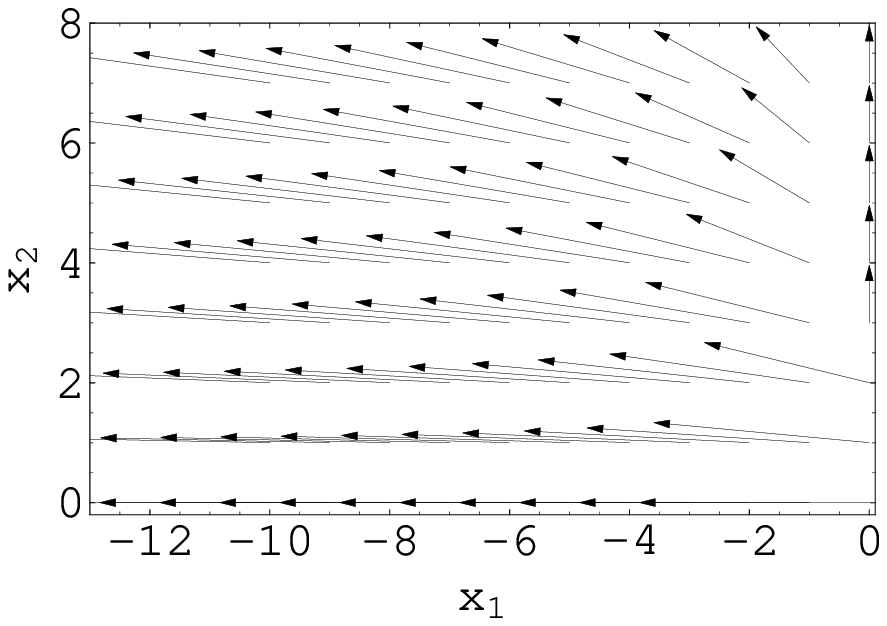}}
\label{image2}
\caption{Astrometric shifts $\Delta \x$ for an SIS perturber in a
 macrolens halo for $\beta=0$ (\ti{left}) and those for 
an SIS perturber in the foreground of a
 macrolens for $\beta=0.5$ (\ti{right}) with respect to images for 
an unperturbed macrolens in the image plane $(x_1,x_2)$ 
aligned to the shear. Both horizontal and vertical axes are
normalized by the Einstein radius $\theta_{Ep}$ of an SIS perturber. 
We adopt a convergence $\kappa=0.38$ and a shear $\gamma=0.47$ 
for the macrolens.}
\end{figure*}

\begin{figure*}
\centerline{\includegraphics[width=8cm]{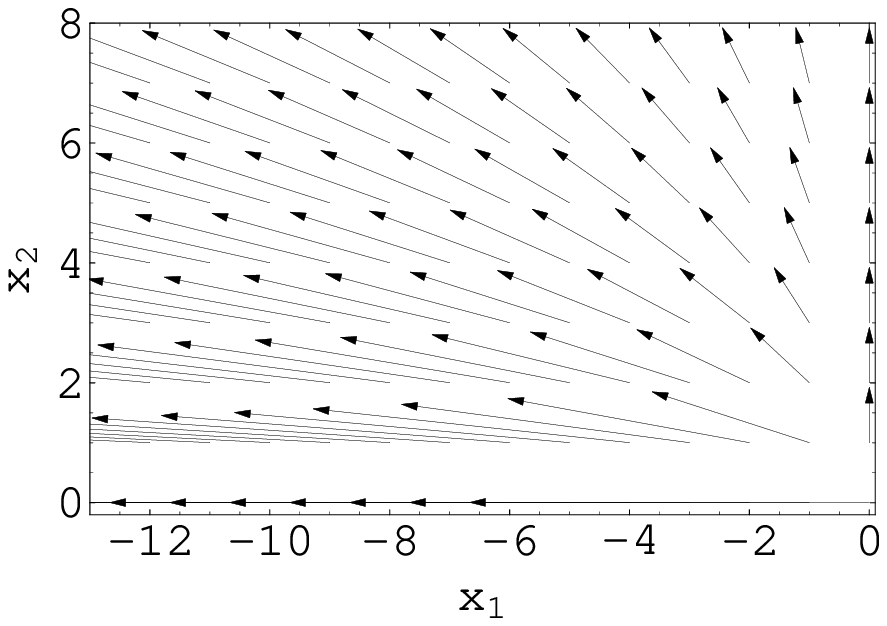}
            \includegraphics[width=8cm]{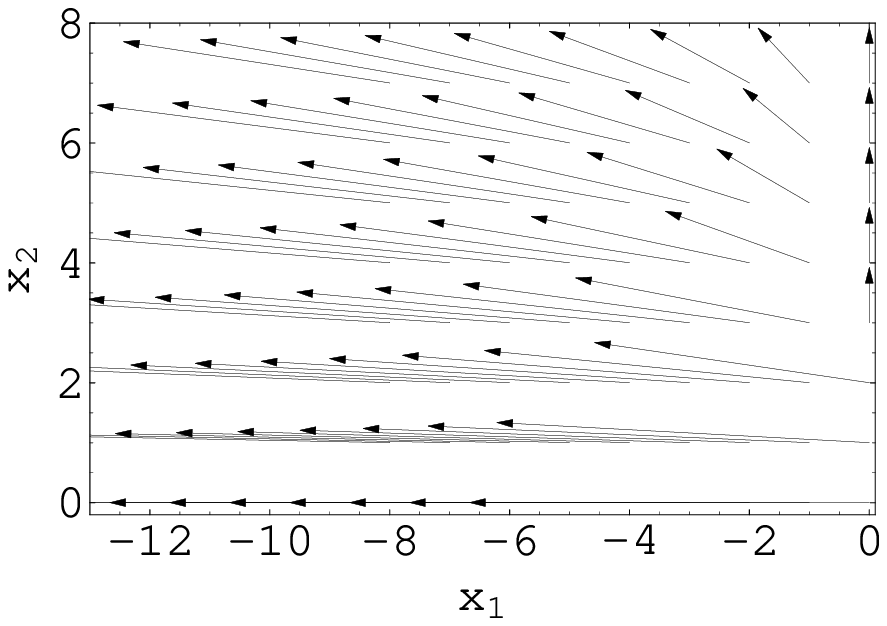}}
\label{image2}
\caption{Astrometric shifts $\Delta \x$ for an SIS perturber in the
 background of a macrolens halo for $\beta=0.9$ (\ti{left}) and 
$\beta=0.5$ (\ti{right}) with respect to images for an unperturbed
macrolens model. The other parameters are the
same as those in Figure 2.}
\end{figure*}

\begin{figure*}[t]
\centerline{
\includegraphics[scale=0.7]{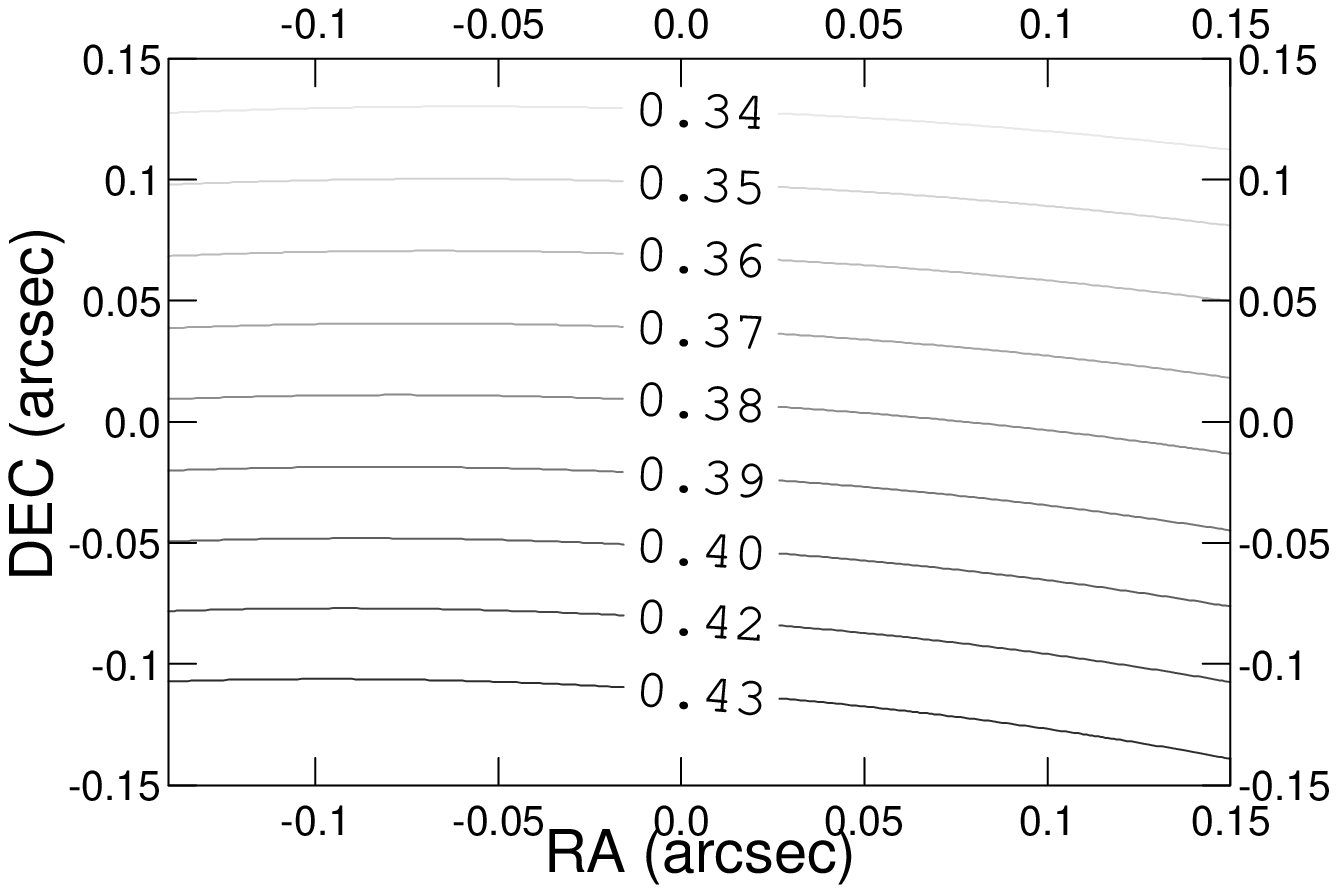}}
\centerline{
\includegraphics[scale=0.7]{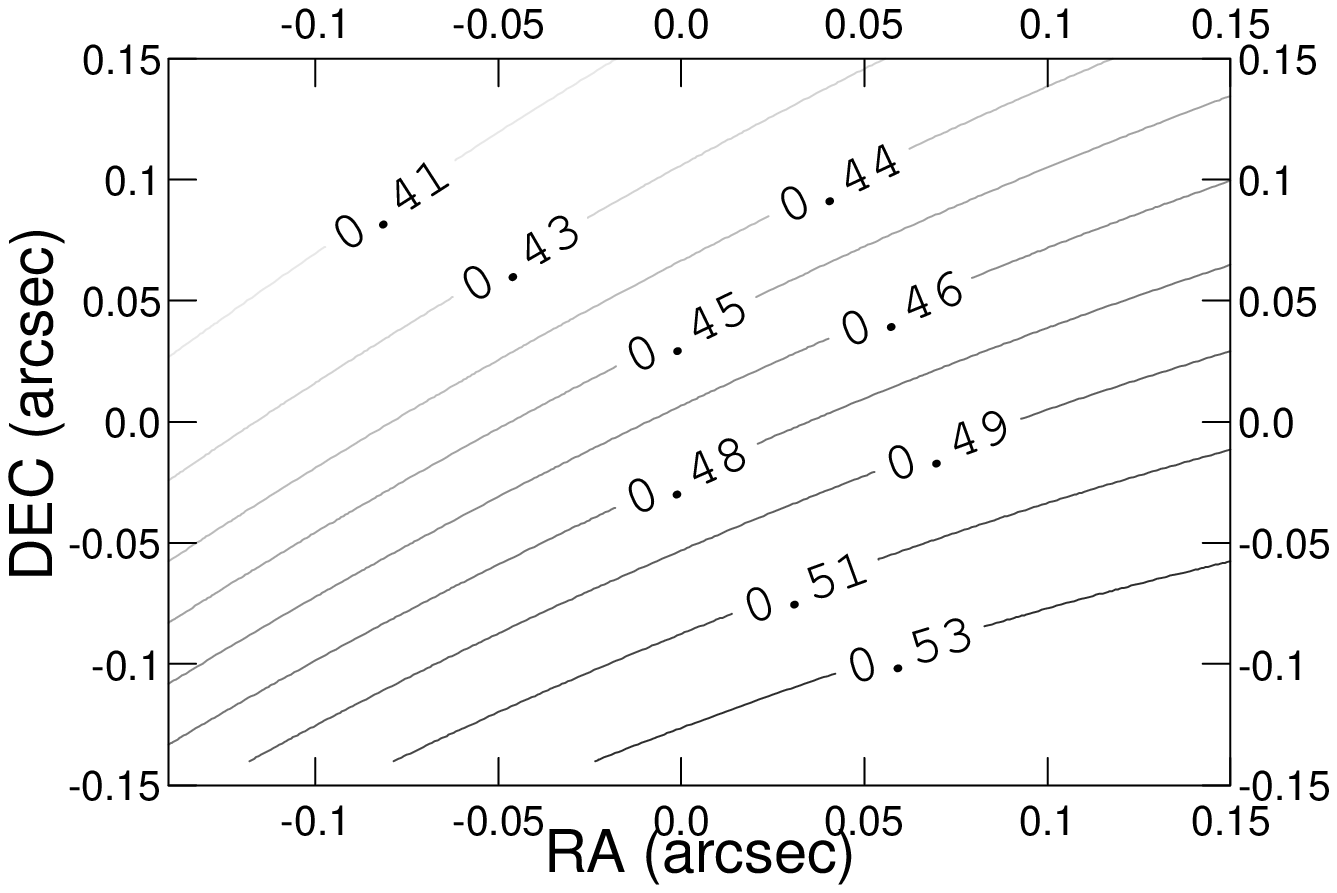}}
\label{image4}
\caption{Contour plots of $\kappa$ (\ti{top}) 
and $\gamma$ (\ti{bottom}) for our SIE model of B1422+231 in the 
neighborhood of an SIS perturber (the position is plotted 
in the top right panel in Figure 1) that is centered at 
the coordinate center.}
\end{figure*}

\section{Summary and Discussion}
In this paper, we have proposed a new 
method to realize three-dimensional mapping of 
CDM substructures in extragalactic halos.
We have shown that at submillimeter wavelengths, a
measurement of astrometric shifts of perturbed 
multiple images
with respect to unperturbed images can 
break the degeneracy between subhalo mass and position
in the line-of-sight to the image 
if resolved at the scale of an Einstein radius of the perturber.
Furthermore, assuming that the tidal radius is approximately equal to
the subhalo size, the distance from the 
center of the macrolens halo 
can be determined from the size of the region within an image 
outside which the astrometric shift is suppressed.

Although we have considered a very simple model, the essential features
of astrometric shifts will not be dramatically altered
even if we consider a more complicated model such as a macrolens perturbed by
a number of non-circular subhalos and/or subhalos having a so-called
NFW profile (Navarro, Frenk \& White 1996).
This is because (1) 
the astrometric shifts caused by a significant number of smaller mass
subhalos are canceled out on larger scales 
if they are not clustered on that scale.
However, several massive perturbers (often not within 
the image) may distort the images so that 
the accuracy in measuring the distance to a single perturber along 
the line-of-sight may become significantly worse.
(2) Deviation from a circular symmetry or from an 
SIS mass profile can affect the astrometric
shifts significantly. For instance, if the substructures 
had an NFW or some other less-cuspy inner profile, the simulated 
dipole pattern inside the Einstein radius would be less
prominent (Inoue \& Chiba 2004). 
However, the generic feature of substructure lensing, i.e., 
its dependence on the parity of a macrolensed image outside 
the Einstein radius of a perturber, will probably be unaltered;
if the outer density profile of a substructure falls off as fast as
an SIS with increasing radius, then we expect similar astrometric
shifts to those for an SIS. Even if the outer density profile is
shallower than an SIS, we can make a distinction between models with
different density profiles and ellipticities from astrometric shifts
of a number of internal positions within the source, if the source has a 
substructure.

We would like to thank the anonymous referee for valuable comments.
This work has been supported in Part by a Grant-in-Aid for Scientific 
Research (15540241) of the Ministry of Education, Culture, Sports, 
Science and Technology in Japan.

\end{document}